\documentclass[amsmath,amssymb,aps, prb, superscriptaddress, 10pt]{revtex4-1}

\usepackage{graphicx}
\usepackage{epstopdf}
\usepackage{color}
\usepackage{bm}
\usepackage{natbib}
\usepackage{multirow}
\newcommand{\im}{\mathrm i}
\newcommand{\ve}{\varepsilon}
\newcommand{\vn}[1][]{\bm{n}_{#1}}
\newcommand{\vs}{\bm{\sigma}}
\newcommand{\dvs}{\cdot\vs}

\graphicspath{ {figures/} }

\begin{document}
\title{Concatenated Composite Pulses \\ Applied to 
Liquid-State Nuclear Magnetic Resonance Spectroscopy}

\author{Masamitsu Bando}
    \affiliation{{Kindai University Technical College, 7-1 Kasugaoaka, Nabari, Mie 518-0459, Japan}}
\author{Tsubasa Ichikawa}
    \affiliation{Department of Physics, Gakushuin University, 1-5-1 Mejiro, Toshima-ku, Tokyo 171-8588, Japan}
\author{Yasushi Kondo}
    \email{ykondo@kindai.ac.jp}
    \affiliation{{Research Center for Quantum Computing, Interdisciplinary Graduate School of Science and Engineering, Kindai University, 3-4-1 Kowakae, Higashi-Osaka, Osaka 577-8502, Japan}}
    \affiliation{Department of Physics, Kindai University, 3-4-1 Kowakae, Higashi-Osaka, Osaka 577-8502, Japan}
    \affiliation{Science and Technology Research Institute, Kindai University, 3-4-1 Kowakae, Higashi-Osaka, Osaka 577-8502, Japan}
\author{Nobuaki Nemoto}
    \affiliation{JEOL RESONANCE Inc., 3-1-2 Musashino, Akishima, Tokyo 196-8558, Japan}
\author{Mikio Nakahara}
    \email{nakahara@shu.edu.cn}
    \affiliation{Department of Mathematics, Shanghai University, 99 Shangda Road, Shanghai 200444, P. R. China}
\author{Yutaka Shikano}
    \email{yutaka.shikano@keio.jp}
    \affiliation{Quantum Computing Center, Keio University, 3-14-1 Hiyoshi, Yokohama, Kanagawa 223-8522, Japan}
    \affiliation{Institute for Quantum Studies, Chapman University, 1 University Dr., Orange, CA 92866, USA}
    \affiliation{Research Center of Integrative Molecular Systems (CIMoS), Institute for Molecular Science, National Institutes of Natural Sciences, 38 Nishigo-Naka, Myodaiji, Okazaki 444-8585, Japan}
    \affiliation{Materials and Structures Laboratory, Tokyo Institute of Technology, 4259 Nagatsuta, Midori, Yokohama, Kanagawa 226-8503, Japan}
    \affiliation{Research Center for Advanced Science and Technology (RCAST), The University of Tokyo, 4-6-1 Komaba, Meguro, Tokyo 153-8904, Japan}

\begin{abstract}
The error-robust and short composite operations named ConCatenated Composite Pulses (CCCPs), developed as high-precision unitary operations in quantum information processing (QIP), are derived from composite pulses widely employed in nuclear magnetic resonance (NMR). CCCPs simultaneously compensate for two types of systematic errors, which was not possible with the known composite pulses in NMR. Our experiments demonstrate that CCCPs are powerful and versatile tools not only in QIP but also in NMR.
\end{abstract}
\maketitle

\section*{Introduction}
\label{sec:intro}
Nuclear magnetic resonance (NMR) is widely used for chemical analysis of various molecules by pharmaceutical companies \citep{TDWC99} owing to highly developed NMR techniques \citep{levitt}. Some of these advanced techniques in NMR have been transferred to quantum information processing (QIP) \citep{JAJ11} because NMR manipulations are regarded as controlling and measuring quantum objects, called spins. We have been working on transferring one of the existing NMR techniques, a composite pulse~\citep{CC85,RT85,MHL86,MHL96} that realises a reliable single spin rotation with erroneous pulses, to QIP. There are two types of composite pulses in NMR: One compensates for pulse-length errors (PLEs), whereas the other compensates for off-resonance errors (OREs). PLEs correspond to rotation angle errors in the dynamics 
of the qubit on a Bloch sphere, and OREs to rotation axis errors. We have successfully 
developed an error-robust and short-pulse-length composite operation (pulses), named 
ConCatenated Composite Pulses (CCCPs), by combining the above-mentioned two types of 
composite pulses in an effort to develop high-precision unitary operations \citep{ota, kindai} 
for QIP. This third type of composite pulse simultaneously compensate for the two types of 
errors (PLEs and OREs in NMR) at the cost of operation time, which was not possible with 
the known composite pulses in NMR.

The purpose of this paper is to feedback our achievement for QIP to NMR. CCCPs are able 
to lead significant signal strength improvement without any changes in the hardware settings.

Let us briefly review the principle of composite pulses compensating PLEs or OREs 
in NMR~\citep{MHL86}. Throughout this paper, the system is a nucleus with spin $1/2$ 
(in short, a {\it spin}) in a static magnetic field along the $z$-axis. An ideal rotation 
operation of the spin without errors is given as
\begin{equation}
 R(\theta,\phi) = \exp[-\im\theta\vn(\phi)\dvs/2], 
 \label{eq:pulse_no_error}
\end{equation}
where $\theta$ is the rotation angle, $\vn(\phi)=(\cos\phi, \sin\phi,0)$ is the 
rotation axis in the $xy$-plane, and $\vs = (\sigma_x, \sigma_y, \sigma_z)$ is 
the Pauli matrices. 
This rotation may be realized by a radio-frequency pulse in NMR, the frequency of which 
is the same as the Larmor frequency of the spin. 

We consider a realistic pulse in which a PLE and/or an ORE are present. 
The first-order terms of the errors are discussed since we are interested in the cases 
where the errors are small. The rotation operator $R'_\ve(\theta,\phi)$ associated 
with a pulse under a PLE  is given as
\begin{eqnarray}
 R'_\ve(\theta,\phi) = \exp[-\im(1+\ve)\theta\vn\dvs/2] 
 = R(\theta,\phi) -\im\ve\theta(\vn\dvs)R(\theta,\phi)/2,
 \label{eq:pulse-ple}
\end{eqnarray}
where $\ve$ is the strength of the PLE, which is unknown --- but constant and small. 
Higher-order terms 
beyond the first order in $\ve$ are suppressed in the second equality. 
This type of error often cannot be avoided 
because of inhomogeneity in the $B_1$ field~\citep{Elham}. By contrast, the rotation operator $ R'_f(\theta,\phi)$ 
associated with a pulse under an ORE is given as
\begin{eqnarray}
 R'_f(\theta,\phi) = \exp[-\im\theta(\vn\dvs + f\sigma_z)/2] 
 = R(\theta,\phi) -\im f\sin(\theta/2)\sigma_z,
  \label{eq:pulse-ore}
\end{eqnarray}
where $f$ is the strength of the ORE. OREs are caused whenever the Larmor frequency 
of the spin is not the same as the transmitter frequency. Therefore, OREs cannot be
avoided in NMR measurements because of the chemical shifts of spins.
As with a PLE, $f$ is unknown --- but constant and small. Therefore, 
when both a PLE and an ORE are present, the rotation associated with a pulse is given as
\begin{eqnarray}
\label{eq:pulse-pleore}
 R'(\theta,\phi) 
&=& \exp[-\im(1+\ve)\theta(\vn\dvs + f\sigma_z)/2] \\
&=& R(\theta,\phi) -\im\ve\theta(\vn\dvs)R(\theta,\phi)/2-\im
  f\sin(\theta/2)\sigma_z. \nonumber 
\end{eqnarray}
The second line is an approximation when both $\ve$ and $f$ are small. 

The NMR community has developed a technique to overcome PLEs or OREs by combining 
several pulses~\citep{JAJ11,MHL86,MHL96}. 
Given a target rotation $R(\theta, \phi)$, we can find an equivalent rotation 
sequence that is equal to the target 
$R(\theta,\phi)$ in a case without errors, as follows:
\begin{equation}
R(\theta_N,\phi_N)R(\theta_{N-1},\phi_{N-1})\cdots R(\theta_1,\phi_1)= R(\theta,\phi).
\label{rs}
\end{equation}
Here, $R(\theta_i,\phi_i)$ is the $i$-th rotation associated with the $i$-th pulse, 
and $N$ denotes the number of pulses. 
The point of the decomposition (\ref{rs}) is 
\begin{equation*}
R'(\theta_N,\phi_N)R'(\theta_{N-1},\phi_{N-1})\cdots R'(\theta_1,\phi_1) \ne R'(\theta,\phi)
\label{rs_error}
\end{equation*}
if a PLE and/or an ORE exist. This non-equality is caused by the non-commutativity 
among $R(\theta_i,\phi_i)$. Therefore, 
by appropriately tuning the parameters $\{\theta_i, \phi_i\}_{i=1}^N$ in Eq.~(\ref{rs}), 
we may be able to obtain a sequence that 
(i) virtually works as the target $R(\theta, \phi)$ when there are no errors, and 
(ii) is less sensitive to the systematic errors.  Indeed, various pulse sequences 
have been designed~\citep{CC85,RT85,MHL86,SW94,HKC03,KRB04,WGA07} 
in such a way that Eq.~(\ref{rs}) has no first-order terms of errors 
if only one of $\ve$ and $f$ exists~\citep{MHL86}. 
We state that such a pulse sequence without the first-order term of $\ve$ ($f$) is 
{\it robust} against PLEs (OREs). 

We now present two typical composite pulses that are robust against either PLEs or 
OREs: Broad Band 1 (BB1)~\citep{SW94}, and Compensation for Off-Resonance 
with a Pulse SEquence (CORPSE)~\citep{HKC03}. See more details in Methods. BB1 is 
designed in order to compensate for a PLE and behaves  as
\begin{equation}
 R'_{\rm BB1}(\theta,\phi)
 = R(\theta,\phi) -\im f\sin(\theta/2)\sigma_z,
 \label{eq:bb11}
\end{equation}
under both a PLE and an ORE. BB1 filters out the PLE but leaves the ORE unchanged, 
which  we call the residual error preserving property (REPP) with respect to ORE. 
In contrast to BB1, CORPSE is a composite pulse robust against OREs and behaves as
\begin{equation}
 R'_{\rm CORPSE}(\theta,\phi) = R(\theta,\phi) -\im\ve(\vn\dvs)R(\theta,\phi)/2.
\end{equation}
Thus, CORPSE possesses REPP with respect to PLE. Not all composite pulses have REPP, 
which was not known before Ref.~\citenum{MB13}. 

We show how to design a CCCP that compensates for both a PLE and an ORE simultaneously 
by taking advantage of REPP, with BB1 and CORPSE as an example~\citep{TI11,MB13}. 
BB1 is robust against PLEs, and CORPSE is robust against  OREs and has the REPP with respect to 
the PLE. Therefore, we replace all pulses in BB1 with CORPSE. This CCCP is called 
CORPSE-in-BB1, or CinBB in short. 
The number of pulses in CinBB is $4\times 3 = 12$. The number of pulses in CinBB can 
be further reduced to $N=6$, and the resulting CCCP is called the reduced CinBB 
(R-CinBB). See Methods and Ref.~\citenum{MB13} for further details. 
Another interesting approach to tackle both PLEs and OREs was discussed 
by Jones~\citep{JAJ13}, in which 
composite pulses were designed to compensate for higher-order 
error terms of both PLEs and OREs simultaneously. The rotation angle $\theta$ is, 
however, fixed to $\pi$ in these composite pulses. See the review 
 by Merrill and Brown~\citep{JTM12} on composite pulses including CCCPs. 

\begin{figure}[ht]
\begin{center}
 \includegraphics[width=7cm]{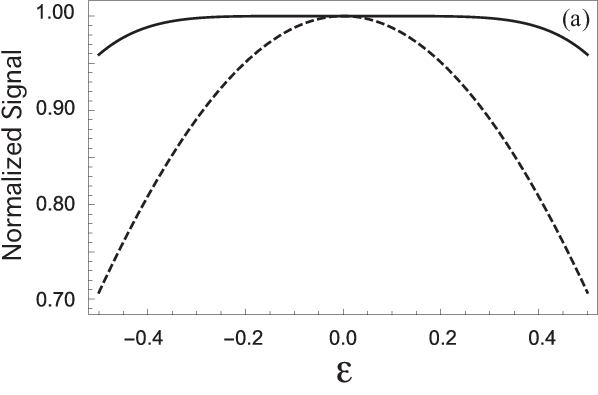} \ \ \
 \includegraphics[width=7cm]{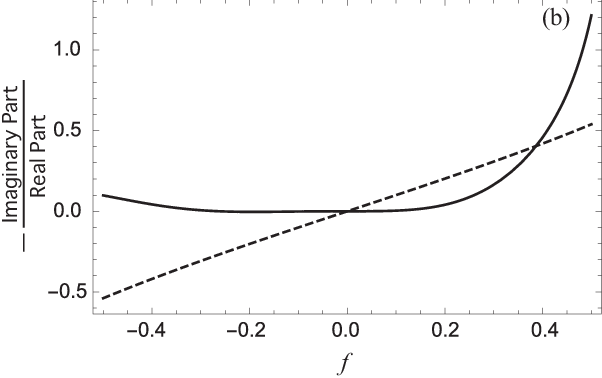}
 \caption{The effect of error size on signal. (a) Normalized signal amplitude after a $\pi/2$-pulse.  
The dashed (solid) curve  is the outcome of a single square (R-CinBB) pulse 
as a function of $\ve$
 (the strength of the pulse length error). 
 (b) ${\rm -\frac{Imaginary \, Part}{Real \, Part} }$, 
a measure of  the direction error from the $x$-axis (the direction of 
the magnetization without errors),  is plotted as a function of $f$
(the strength of the off-resonance error).  
The calculations with Eq.~(\ref{eq:pulse-pleore}) are performed without approximation because $|\ve| \sim 0.5$ 
and $|f| \sim 0.5$ cannot be regarded as small.}
\label{fig:cccp}
\end{center}
\end{figure}

The signal after a single square $\pi/2$-pulse is shown as the dashed lines in 
Fig.~\ref{fig:cccp}. This single square pulse has a constant $B_1$ 
during the period of $\tau_p$ and $B_1 \tau_p $ is $\pi/2$. Its rotation axis 
in the Bloch sphere is, here,  the $y$-axis, and thus the magnetization after the 
$\pi/2$-pulse is in parallel to the $x$-axis if there are no errors. 
Figure~\ref{fig:cccp}(a) shows the normalized signal as a function of 
$\ve$ (the dotted curve); $\ve$ as small as $\ve = 0.1$ leads to a significant 
signal reduction. Figure~\ref{fig:cccp}(b) shows that the magnetization 
after the single square $\pi/2$-pulse deviates from the $x$-axis and its deviation appears 
to be proportional to $f$.
Then, let us consider the signal after the R-CinBB $\pi/2$-pulse which consists of 
six square pulses (see Methods for details). 
The solid line in Fig.~\ref{fig:cccp}(a) shows that one obtains
a larger signal for a wide range of $\ve$ with the R-CinBB $\pi/2$-pulse than 
with the single square $\pi/2$-pulse. 
On the other hand, the solid line in Fig.~\ref{fig:cccp}(b) shows 
that the magnetization after the R-CinBB $\pi/2$-pulse is close to the $x$-axis 
for a wider range of $f$ than after the single square $\pi/2$-pulse. 

\section*{Results}
\subsection*{Simulations of NMR experiments}
\label{A_sim}
Let us take into account a non-unitary time development caused by a spin--spin relaxation 
with a characteristic time $T_2$ in simulating NMR experiments. We introduce this effect 
as a phase flip channel~\citep{Nielsen}. In the case of single-spin experiments, 
\begin{align}
 \rho(t+\Delta ) &= p_{ss}(\Delta ) \rho(t) + \left(1-p_{ss}(\Delta )\right)
{\rm Ad} \left(\sigma_z, \rho(t)\right), 
\end{align}
where $p_{ss}(\Delta ) = (1 + \exp(-\Delta /T_2))/2 \approx 1 - \Delta/2T_2$ and
${\rm Ad}(\xi, \rho) = \xi^\dagger \rho \xi$ with an arbitrary unitary operator 
$\xi$. $\Delta$ is a small time interval. The subscript $ss$ denotes ``spin-spin''. 

The time evolution during a pulse is simulated as follows:
\begin{align}
\label{eq_FID}
\tilde{\rho}(t+\tau_p) &= p_{ss}(\tau_p)\rho(t) + \left(1-p_{ss}(\tau_p )\right)
{\rm Ad} \left(\sigma_z, \rho(t)\right), \nonumber \\
\rho(t+\tau_p) &= {\rm Ad}(U_{\rm pulse}, \tilde{\rho}(t+\tau_p)), 
\end{align}
where $U_{\rm pulse}$ is a unitary operation generated by the pulse. Note that $\tau_p$ is 
the total pulse duration and is assumed to be small. Therefore, we employ the Suzuki-Trotter 
formula, which ensures the decomposition of the dynamical evolution into the form of pure 
relaxation process followed by the application of the composite pulse~\citep{SZ}.

\begin{figure}[ht]
 \center
 \includegraphics[width=0.42\linewidth,clip]{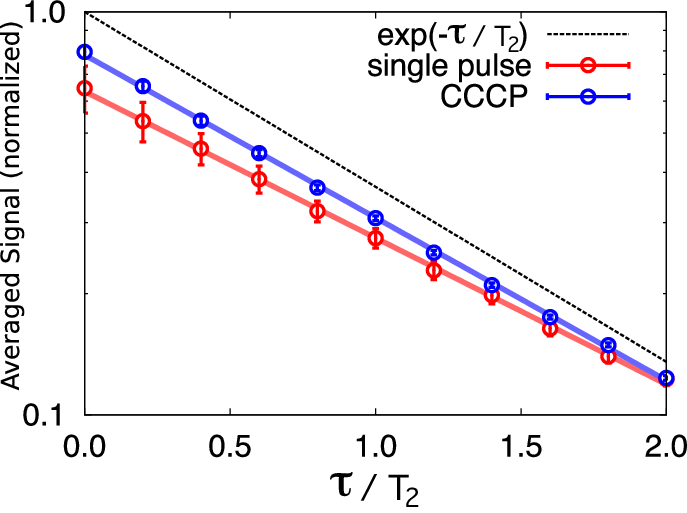}
 \caption{Semi-log plot of echo signals with two square (red line) and two 
R-CinBB (blue line) pulses as functions of the waiting time $\tau$. 
The black dashed line is an  error-free case. Error bars represent the fluctuation 
of the signal strength. The figure shows that the R-CinBB pulses suppress the fluctuation 
of the echo signals. }
 \label{fig:cccp-edecay}
\end{figure}

\begin{figure}[ht]
\begin{center}
 \includegraphics[width=0.8\linewidth]{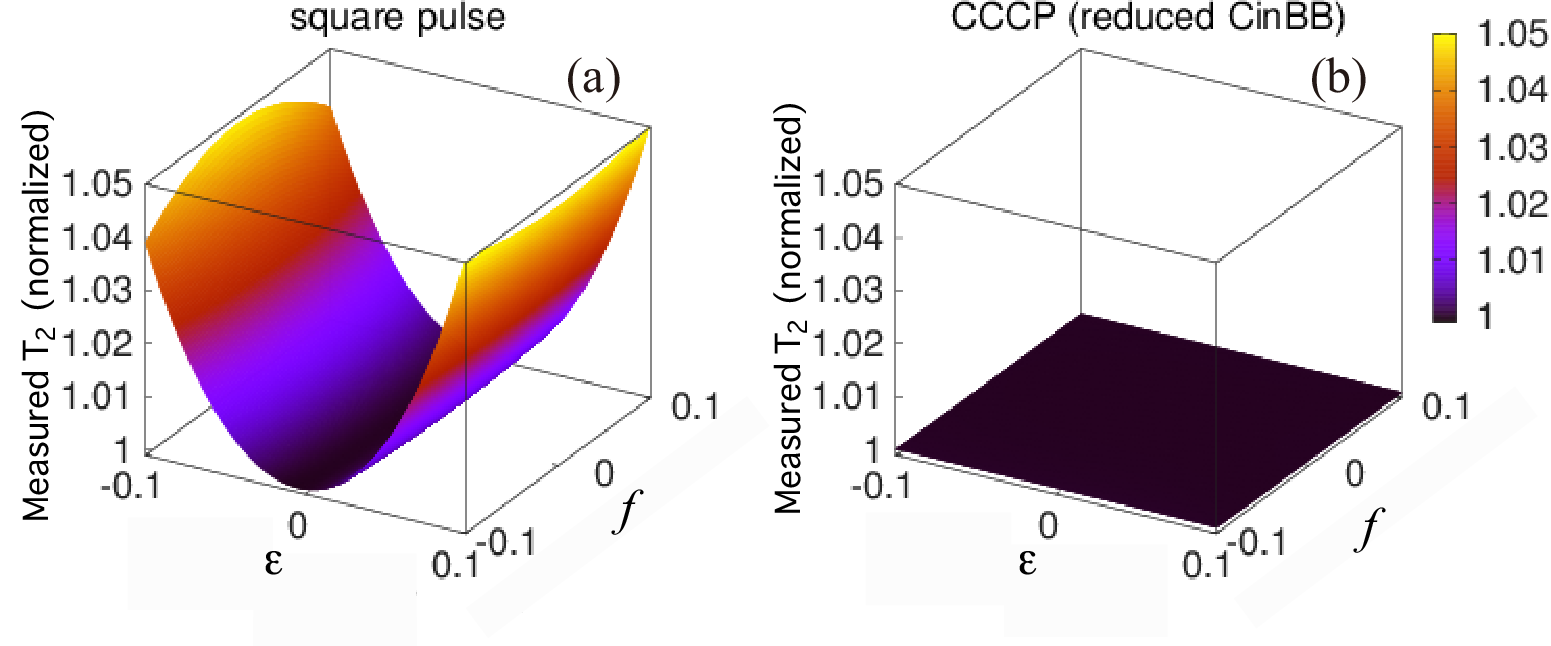}
\end{center}
 \caption{
Measured $T_2$ 
as a function of a PLE and an ORE for (a) square pulses and (b) 
R-CinBB pulses. 
 The Hahn echo experiments with R-CinBB pulses lead to 
the correct $T_2$, 
even in erroneous cases. Measured $T_2$'s are normalized by the true 
$T_2$.}
 \label{fig:cccp-t2}
\end{figure}

We examine Hahn echo experiments~\citep{levitt} with two pulses which are affected 
by fluctuating PLEs and OREs. Their means are $\bar{\ve} = \bar{f} = 0.1$  
and their standard deviations are both $0.08$.  Although these values may be unreasonably 
large for modern NMR spectrometers, simulated results show that the echo signals with 
R-CinBB pulses do not fluctuate, as shown in Fig.~\ref{fig:cccp-edecay}.  
Simulations of the Hahn echo experiments as a function of the error strengths are summarized 
in Fig.~\ref{fig:cccp-t2}. The Hahn echo experiments with two single square pulses 
(Fig.~\ref{fig:cccp-t2}a) are strongly affected by PLEs, whereas those with R-CinBB pulses 
are robust against  these errors (Fig.~\ref{fig:cccp-t2}b). It turns out that a composite pulse robust against  PLEs is 
sufficient for obtaining a correct $T_2$ even when both PLEs and OREs are present. 

Let us examine two-dimensional (2D) shift-COrrelation SpectroscopY (COSY) experiments, 
one of the most important NMR measurement methods~\citep{TDWC99}, with two interacting 
spins. The interaction is a scalar coupling in a weak coupling limit~\citep{levitt}.  
The simulations during the evolution and detection periods~\citep{TDWC99} are done as follows:
\begin{align}
\label{Eq_COSY}
\tilde{\rho}(t+\delta)&= 
\left(1- \frac{\delta}{2T_{2,1}} -  \frac{\delta}{2T_{2,2}} \right)\rho(t) 
+ \frac{\delta}{2T_{2,1}}{\rm Ad}(\sigma_z \otimes \sigma_0, \rho(t)) 
+ \frac{\delta}{2T_{2,2}}{\rm Ad}(\sigma_0 \otimes \sigma_z, \rho(t)),  \nonumber \\
\rho(t+\delta) &= {\rm Ad} \left(\exp \left(-J\delta \frac{\sigma_z \otimes \sigma_z}{4} 
\right), 
\tilde{\rho}(t+\delta) \right),
\end{align}
where $T_{2,i}$ is the spin-spin relaxation time of the $i$-th spin. Eq.~\eqref{Eq_COSY} 
is again, similarly to Eq.~\eqref{eq_FID}, based on the Suzuki-Trotter formula~\citep{SZ}: 
The first equation in Eq.~\eqref{Eq_COSY} describes the spin-spin 
relaxation channel and the second one is the time development generated by the 
spin-spin interaction.

\begin{figure}[ht]
\begin{center}
 \includegraphics[width=7cm]{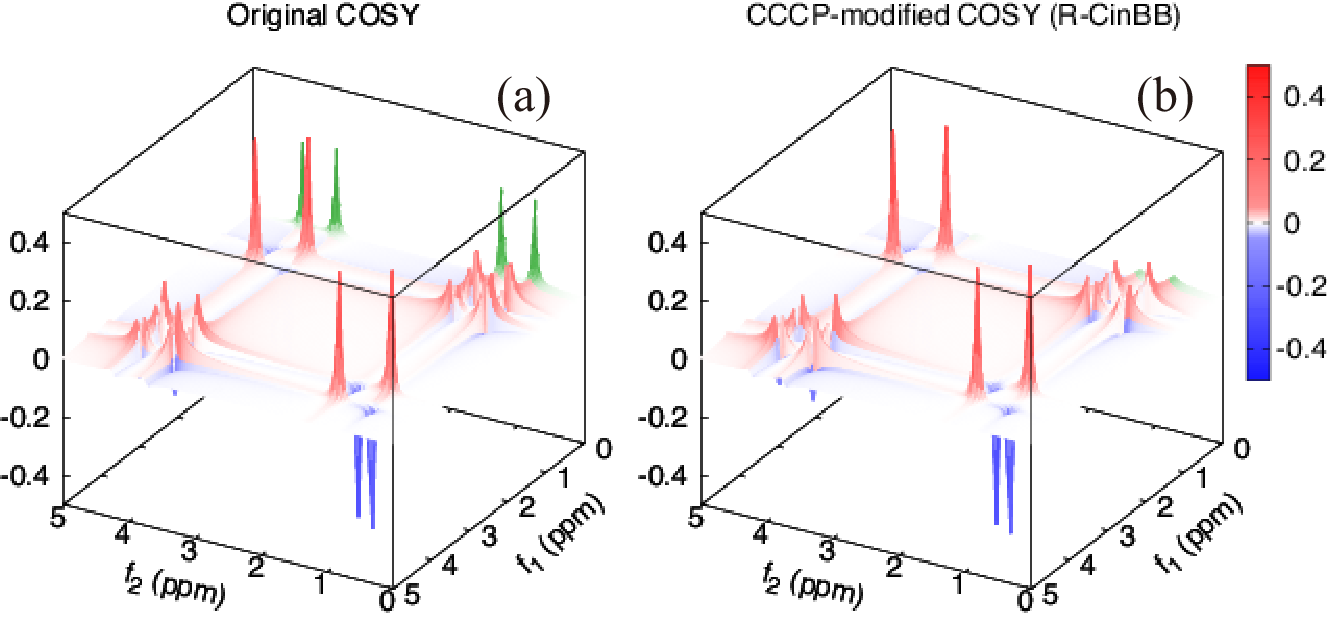}
\end{center}
 \caption{Simulation of COSY experiments of a two-spin molecule with (a) two  
single square and (b) R-CinBB pulses, where $\ve=f=0.1$, $J=0.5$~ppm, 
and chemical shifts are $1$ and $4$~ppm.  Spurious axial (green) peaks are observed 
in the simulation with square pulses, whereas no such peaks are observed in that
with R-CinBB pulses.}
 \label{fig:cosy}
\end{figure}

$\rho(t+n \delta)$ can be obtained by iterating the above operations $n$ times. 
Note that $p_{ss}(\delta) \approx 1 - \delta/(2 T_{2,i}) $ for the $i$-th spin 
because $\delta$ is sufficiently small compared to $T_{2,i}$. 
During a pulse, the time development is simulated similarly to the case 
of single-spin experiments. 
Simulations are summarized in Fig~\ref{fig:cosy} in the case that the chemical 
shifts of these spins are $1$ and $4$~ppm and $J = 0.5$~ppm. In COSY experiments, 
spurious peaks called axial peaks sometimes appear owing to the inaccuracy of 
the first pulse~\citep{TDWC99}. 
We are able to reproduce these axial peaks in the simulation of the COSY experiments 
with two single square pulses ($\ve = f = 0.1$), as shown in 
Fig.~\ref{fig:cosy}a. By contrast, no axial peaks appear in the simulation 
with R-CinBB pulses in Fig.~\ref{fig:cosy}b. 

\subsection*{Experimental demonstration of NMR measurements with CCCPs}
\label{sec:adv}
\begin{figure}[ht]
\begin{center}
\includegraphics[width=7cm]{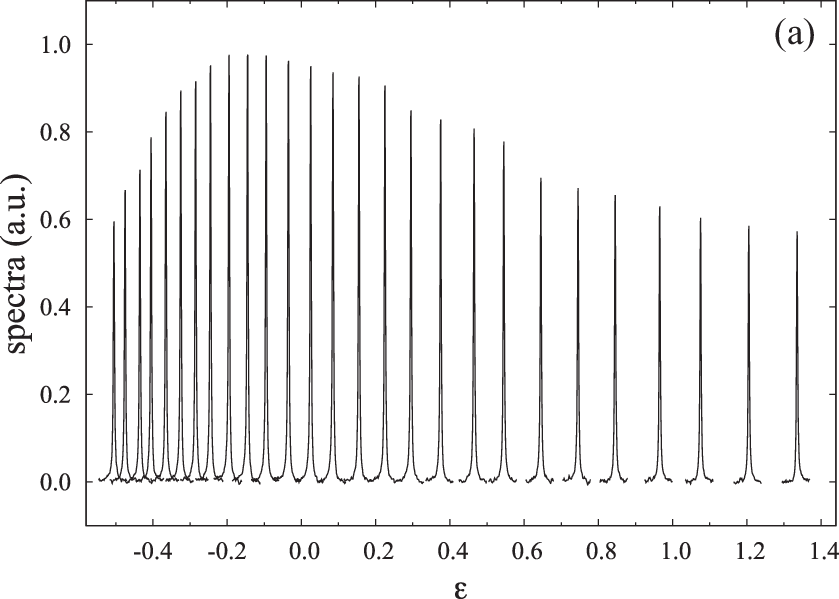}
\includegraphics[width=7cm]{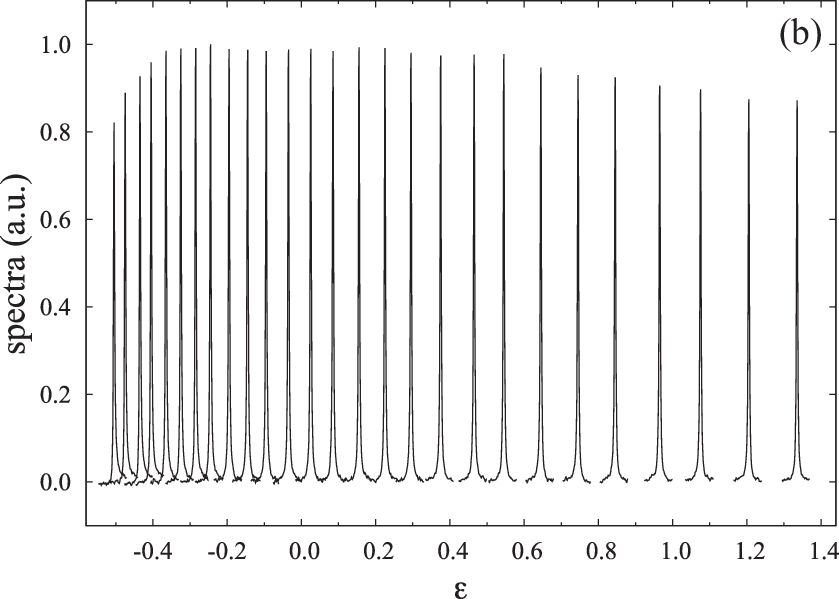}
 \caption{Series of 1D spectra with (a) square $\pi/2$-pulse 
and (b) R-CinBB $\pi/2$-pulse applied as functions of the PLE 
($\ve$ in Eq.~(\ref{eq:pulse-pleore})). The strength of the ORE is 
fixed at $f \sim 0$.}
 \label{fig:JEOL-ple}
\end{center}
\end{figure}

\begin{figure}[t]
\begin{center}
 \includegraphics[scale=0.45]{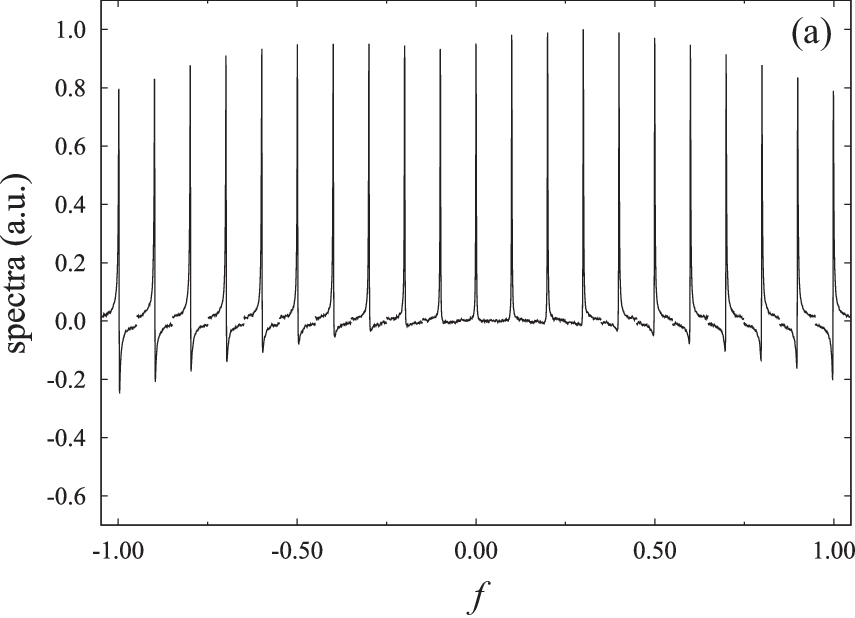}
 \includegraphics[scale=0.45]{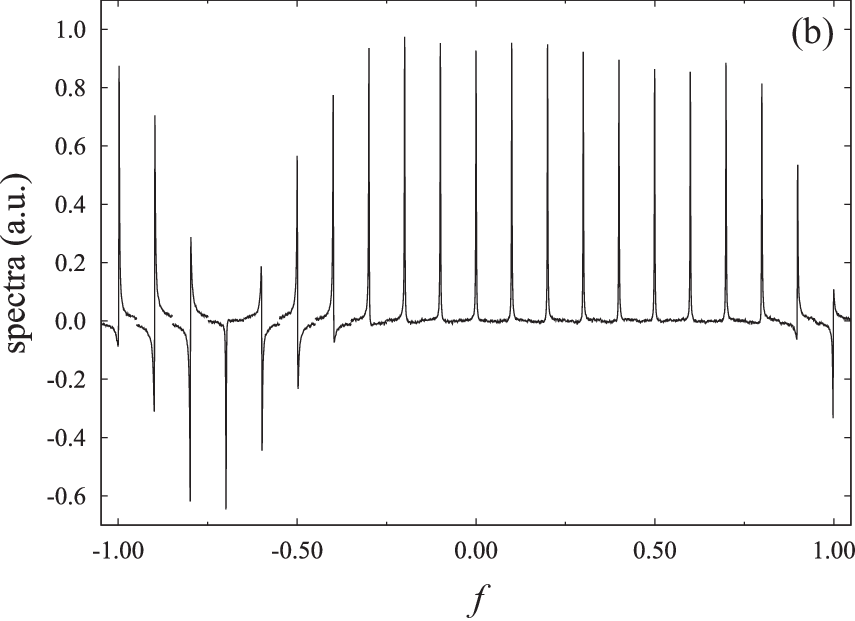}
 \caption{Series of 1D spectra with (a) square $\pi/2$-pulse 
and (b) R-CinBB $\pi/2$-pulse applied as functions of the ORE 
($f$ in Eq.~\ref{eq:pulse-pleore}). The strength of the PLE is 
fixed at $\varepsilon \sim 0$.}
 \label{fig:JEOL-ore}
\end{center}
\end{figure}

The advantages of CCCPs in NMR are demonstrated in the following experiments. 
The single-pulse experiments were carried out using $300$~mM $^{13}$C-labelled 
chloroform in acetone-$d_6$ at 25{}$^\circ$C.  We examined the performance 
of a composite $\pi/2$-pulse applied to $^{13}$C and compared the result 
with that of a single square pulse~\citep{AJS83}. It is clear that the 
R-CinBB pulses are more advantageous than single square pulses 
in terms of PLE, as shown in Fig.~\ref{fig:JEOL-ple}. This is also demonstrated in the corresponding 
numerical calculations, shown in Fig.~\ref{fig:cccp}(a). We also examined the 
R-CinBB pulse in terms of the ORE, as shown in Fig.~\ref{fig:JEOL-ore} 
(see also Fig.~\ref{fig:cccp}(b)).
The R-CinBB composite pulse is clearly more advantageous than the square 
pulse when $-0.3 < f < 0.8$. Although the spectra with the R-CinBB composite 
pulses corresponding to $f < -0.5$ and $1.0 < f$ are more distorted than those 
of the single square pulses, such large $f$'s are not relevant in usual 
experiments. See Ref.~\citenum{MB13} for details. 

\begin{figure}[ht]
\begin{center}
 \includegraphics[scale=0.9]{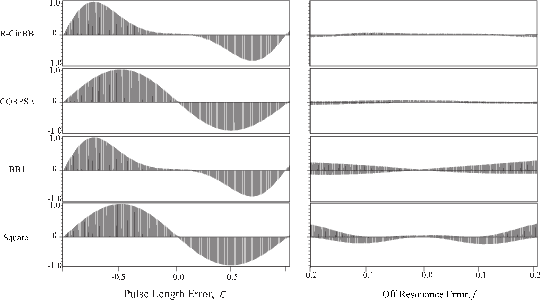}
 \caption{ Series of 1D $^1$H-NMR spectra of 2$\rm \%$ HDO solution measured after two
successive (R-CinBB, CORPSE, BB1, and square) $\pi/2$-pulses. 
All the spectra shown in these eight panels 
are normalized with the signal intensity obtained by a  single square $\pi/2$-pulse 
spectrum (data not shown). 
The single square $\pi/2$-pulse duration is 9.95~$\mu$s. 
Each panel contains 201 1D spectra. 
 \label{fig:hpi-pulses}
}
\end{center}
\end{figure}

The advantage of the R-CinBB $\pi/2$-pulse in NMR is also evaluated, as shown in Fig.~\ref{fig:hpi-pulses}. We applied two successive (R-CinBB, CORPSE, 
BB1, or square) $\pi/2$-pulses to the thermal equilibrium state. 
A pair of successive $\pi/2$-pulses is equivalent to a single $\pi$-pulse 
without errors and should lead to no signal. Therefore, the observed residual 
signals are measures of errors in these pulses. The advantage of 
the R-CinBB $\pi/2$-pulse is clear from the fact that the two successive R-CinBB $\pi/2$-pulses 
lead to  small signals in wider ranges of both PLEs ($\ve$) and OREs ($f$). 

\begin{figure}[ht]
\begin{center}
\includegraphics[scale=0.9]{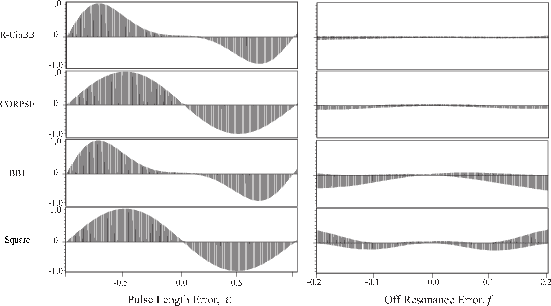}
 \caption{ Series of 1D $^1$H-NMR spectra of 2$\rm \%$ HDO solution measured 
after single (R-CinBB, CORPSE, BB1, and square) $\pi$-pulses. 
All the spectra shown in these eight panels 
are normalized with the signal intensity obtained by the single square 
$\pi/2$-pulse spectrum used in Fig.~\ref{fig:hpi-pulses}. 
Each panel contains 201 1D spectra. 
 \label{fig:pi-pulses}
}
\end{center}
\end{figure}

The advantage of the R-CinBB pulse in NMR is also evaluated in the case of 
$\pi$-pulses, as shown in Fig.~\ref{fig:pi-pulses}. We applied a (R-CinBB, 
CORPSE, BB1, or square) $\pi$-pulse to the thermal equilibrium state. An ideal $\pi$ pulse 
should lead to no signal. The advantage of the $\pi$ R-CinBB pulse is clear from
the fact that the R-CinBB $\pi$-pulses lead to small signals 
in wider ranges of both PLEs ($\ve$) and OREs ($f$). 
These experiments were carried out as in the case of Fig.~\ref{fig:hpi-pulses}. 
It is interesting to note that the behaviours as a function of $\ve$ of CORPSE 
and square pulses 
are identical, which indicates that CORPSE has REPP with 
respect to PLE in the whole range of $\ve$ in Figs.~\ref{fig:hpi-pulses} and 
\ref{fig:pi-pulses}. On the other hand, 
the REPP of BB1 with respect to ORE is only valid for small $|f|$.

\begin{figure}[ht]
\begin{center}
 \includegraphics[scale=0.62]{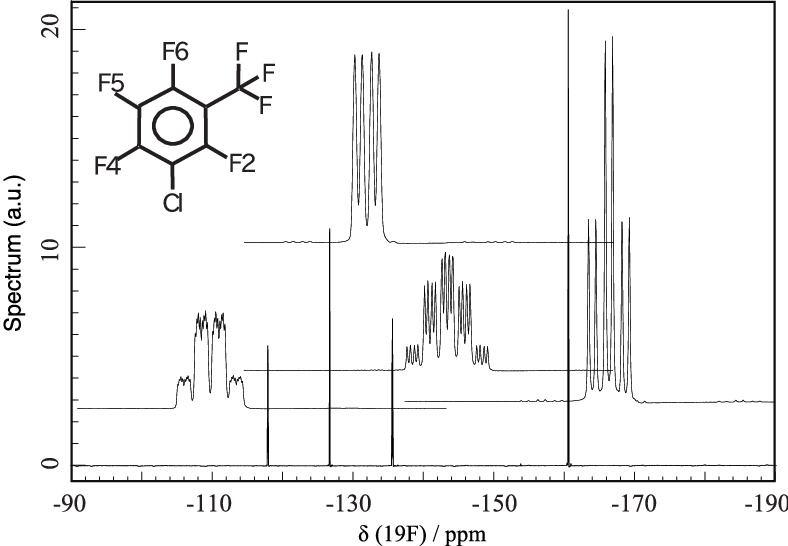}
 \caption{1D $^{19}$F-NMR spectrum of 300 mM 
3-chloro-2,4,5,6-tetrafluoro-benzotrifluoride in benzene-$d_6$. 
Each peak is enlarged to show detailed structures ($1$~ppm width). 
Peak assignments are as follows: 1 ($-118.0$~ppm), 2 ($-126.5$~ppm), 
3 ($-135.5$~ppm), and 4 ($-160.5$~ppm) are identified as F2, F4, F6, and F5, 
respectively. The $^{19}$F signal of the trifluoromethyl group 
is not observed in this frequency region.}
 \label{fig:fbf-1D}
\end{center}
\end{figure}

Next, we performed COSY experiments of 300~mM 
3-chloro-2,4,5,6-tetrafluoro-benzotrifluoride 
in benzene-$d_6$. We utilized $\rm^{19}F$ at 2, 4, 5, and 6 
as the target nuclear spin. $T_1$'s are between $0.6$ and $1.0$~s, 
whereas $T_2$'s are $\sim 0.3$~s.  We chose this molecule 
for the following reasons. First, $^{19}$F signals of the molecule 
are widely spread, as shown in Fig.~\ref{fig:fbf-1D}. Second, 
the spectrum pattern is complex enough to examine the performance 
of the pulses, despite of its simple molecular structure.

\begin{figure}[ht]
\begin{center}
 \includegraphics[width=16cm]{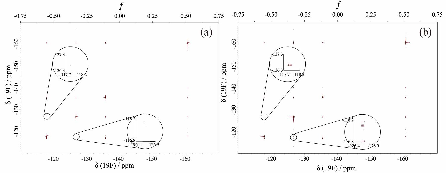}
 \caption{$^{19}$F-$^{19}$F COSY spectra of 300 mM 
{3-chloro-2,4,5,6-tetrafluoro-benzotrifluoride} in benzene-$d_6$ 
obtained with (a) two successive single square pulses and (b) 
R-CinBB pulses. The regions $-118.0$~ppm($f_1$) / $-126.5$~ppm($f_2$) 
and $-126.5$~ppm($f_1$) / $-118.0$~ppm($f_2$) are enlarged.}
\label{fig:JEOL-cosy-all}
\end{center}
\end{figure}

\begin{figure}[ht]
\begin{center}
\includegraphics[width=0.45\linewidth]{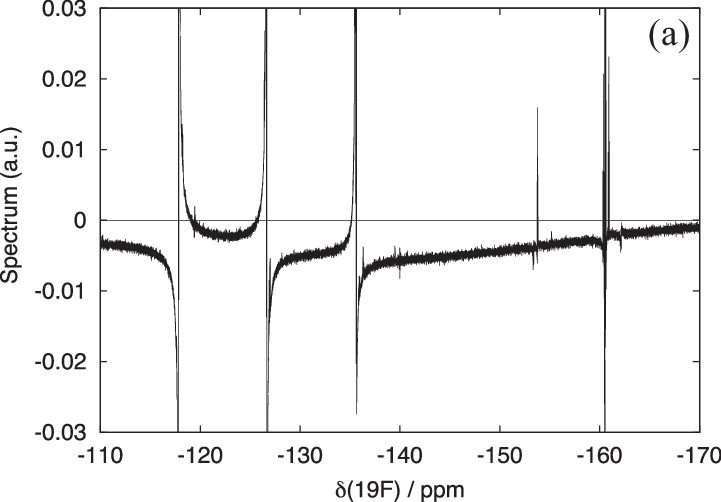}
\includegraphics[width=0.45\linewidth]{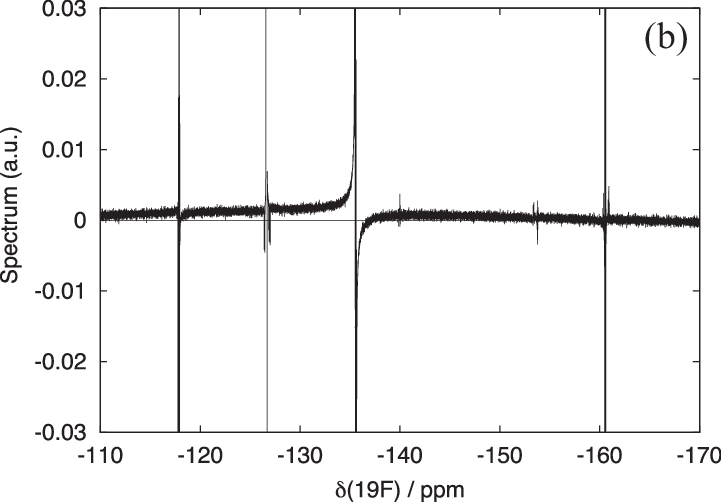}
 \caption{First increments of the $^{19}$F-{}$^{19}$F COSY obtained 
with (a) square pulses and (b) R-CinBB pulses (2D spectra 
are shown in Fig.~\ref{fig:JEOL-cosy-all}). }
 \label{fig:JEOL-cosy-1d}
\end{center}
\end{figure}

\begin{table}[ht]
\centering
\caption{$\pi/2 \ (90^\circ)$ and $\pi \ (180^\circ)$ rotation implemented by BB1, CORPSE, and R-CinBB.}
\begin{tabular}{lcrrrrr}
\hline
&          & \multicolumn{2}{c}{$\theta = 90^\circ$} & \multicolumn{1}{l}{} & \multicolumn{2}{c}{$\theta = 180^\circ$} \\ \cline{3-4}\cline{6-7}
Name                                                                                       & Position & $\theta/$degree     & $\phi/$degree    &                      & $\theta/$degree     & $\phi/$degree     \\ \hline
BB1                                                                                        & 1        & 180.0               & 97.2             &                      & 180.0               & 104.5             \\
(robust for PLEs) & 2        & 360.0               & 291.5            &                      & 360.0               & 313.4             \\
                                                                                           & 3        & 180.0               & 97.2             &                      & 180.0               & 104.5             \\
                                                                                           & 4        & 90.0                & 0.0              &                      & 180.0               & 0.0               \\
                                                                                           &          &                     &                  &                      &                     &                   \\
CORPSE                                                                                     & 1        & 384.3               & 0.0              &                      & 420.0               & 0.0               \\
(robust for OREs)                                                                          & 2        & 318.6               & 180.0            &                      & 300.0               & 180.0             \\
                                                                                           & 3        & 24.3                & 0.0              &                      & 60.0                & 0.0               \\
                                                                                           &          &                     &                  &                      &                     &                   \\
R-CinBB                                                                                     & 1        & 180.0               & 97.2             &                      & 180.0               & 104.5             \\
(robust for both PLEs and OREs) & 2        & 360.0               & 291.5            &                      & 360.0               & 313.4             \\
                                                                                           & 3        & 180.0               & 97.2             &                      & 180.0               & 104.5             \\
                                                                                           & 4        & 384.3               & 0.0              &                      & 420.0               & 0.0               \\
                                                                                           & 5        & 318.6               & 180.0            &                      & 300.0               & 180.0             \\
                                                                                           & 6        & 24.3                & 0.0              &                      & 60.0                & 0.0 \\
\hline            
\end{tabular}
\label{table:pulse}
\end{table}

Here, the pulse duration of a single square $\pi/2$-pulse is 
$12.4$~${\rm \mu}$s, which corresponds to a $B_1$ strength of 
$20$~kHz in frequency. The total duration of the R-CinBB pulse is 
$16.1 \times 12.4 ~$ ${\rm \mu} {\rm s} = 2.00 \times 10^2~{\rm \mu s}$, 
which is almost instantaneous compared with the inverse of the interaction 
strength in frequency. Therefore, the replacement of a square pulse 
by the R-CinBB pulse should not cause problems for most 
applications of liquid state NMR measurements.

Since the $B_1$ strength is 20~kHz, it is comparable to the frequency 
difference between the highest ($-160$~ppm) and the lowest ($-118$~ppm) peaks 
at 11.7 T (488 MHz for $^{19}$F and 500 MHz for $^1$H); see Fig.~\ref{fig:fbf-1D}. 
In the case of square pulses, the correlation peak between $-118.0$~ppm 
($f_1$) and $-126.5$~ppm ($f_2$), and that between $-126.5$~ppm ($f_1$) and 
$-118.0$~ppm ($f_2$), are hardly visible. As shown in Fig.~\ref{fig:JEOL-cosy-all}, 
however, these have much higher intensities in the case of the R-CinBB pulses. 
In addition, the advantage of the R-CinBB pulses is much more clearly demonstrated 
in the one-dimensional (1D) spectra in Fig.~\ref{fig:JEOL-cosy-1d}. 
The phases of peaks obtained with square pulses are highly distorted. This may be 
one of the biggest reasons why the above correlation peaks are almost invisible.

\section*{Discussion}
\label{sec:summary}
Composite pulses have been developed in the NMR community and are widely employed. Our proposed composite operations, CCCPs, directly descend 
from these and have been developed as robust unitary operations for QIP. We feedback 
our achievements to NMR: We applied CCCPs to liquid-state NMR spectroscopy and 
demonstrated improved NMR sensitivity compared to standard 1D and 
2D NMR measurements with square pulses. 
The proposed CCCPs are robust against two systematic errors, the PLE 
and ORE in NMR, at the cost of execution time. 

We demonstrated the advantage of the R-CinBB pulses over 
the BB1, CORPSE, and square pulses in 1D and 2D (COSY) experiments.  
In terms of the compensation of PLEs and OREs, the replacement of single square pulses  
with CCCPs, such as the R-CinBB pulses, should be widely utilized 
in other experiments 
in liquid-state NMR. 
On the other hand, the application to solid-state NMR (SS-NMR) may be limited, 
because shorter pulses are favourable in SS-NMR in general and much longer CCCPs 
might be unacceptable in most cases. 
In the case of SS-NMR, COM-I, II, and III pulses~\citep{com} are often employed 
as wideband (robust against $f$) pulses. These pulses employ only $0^\circ$ and 
$180^\circ$ phase pulses and thus the requirement of the electronics is less demanding
compared with our proposed CCCPs.  We believe, however, that advances in electronics
can now allow use of CCCPs even in SS-NMR experiments. 

CCCPs consist of simple spin rotation pulses and 
thus they are technically easy to implement although they are not optimal 
in terms of quantum control theory~\citep{oc1, oc2, oc3, Koike}.
As mentioned before, composite pulses are widely used in NMR
experiments, and we hope that CCCPs will be employed instead of these 
known composite or single square pulses because of their advantage. We believe that 
CCCPs should be useful for magnetic resonance imaging, too. Furthermore, 
CCCPs might be applied to positron $g/2$ measurements through the use of 
a $^3{\rm He}$-NMR probe~\citep{harvard} in which the inhomogeneity of an excitation field
may be large.  
Also, because the pulse sequence in nonlinear optical 
spectroscopy has been inspired by the NMR pulse techniques~\citep{Mukamel1995}, 
CCCPs may be applicable in such optical systems in order to enhance the accuracy 
of optical spectroscopy.

\section*{Materials and Methods}
\subsection*{BB1}
BB1~\citep{SW94} is an $N=4$ composite pulse robust against PLEs. 
The parameters are as follows:
\begin{align}
 \theta_1 = \theta_3 = \pi,\ \theta_2 = 2\pi,\ \theta_4 = \theta, 
 \phi_1 = \phi_3 = \phi + \arccos[-\theta/(4\pi)],\ 
 \phi_2 = 3\phi_1 - 2\phi,\ \phi_4 = \phi.
\end{align}
BB1 under both a PLE and an ORE results in 
\begin{align}
 R'_{\rm BB1}(\theta,\phi) = R'(\theta,\phi)R'(\pi,\phi_1)R'(2\pi,\phi_2)R'(\pi,\phi_1) 
= R(\theta,\phi) -\im f\sin(\theta/2)\sigma_z.
 \label{eq:bb1}
\end{align}

\subsection*{CORPSE}
CORPSE~\citep{HKC03} is an $N=3$ composite pulse robust against  OREs. Its parameters are
\begin{align}
 \theta_1 = 2n_1\pi + \theta/2 -k,\ 
 \theta_2 = 2n_2\pi -2k,\ 
 \theta_3 = 2n_3\pi + \theta/2 -k,
 \phi_1 = \phi_2 - \pi = \phi_3 = \phi,\ k = \arcsin[\sin(\theta/2)/2],
\end{align}
where $n_1$, $n_2$, and $n_3$ are non-negative integers. 
In particular, when we take $n_1 = n_3 = 0$  and $n_2 = 1$, 
the execution time is minimized. In this case, CORPSE is referred to as short CORPSE. 
Another notable case takes place when $n_1 - n_2 + n_3 = 0$. In this case, with both 
a PLE and an ORE, CORPSE results in 
\begin{align}
 R'_{\rm CORPSE}(\theta,\phi)
 = R'(\theta_3,\phi)R'(\theta_2,\phi+\pi)R'(\theta_1,\phi)
 = R(\theta,\phi) -\im\ve(\vn\dvs)R(\theta,\phi)/2.
\end{align}

\subsection*{Reduced CORPSE in BB1}
R-CinBB~\citep{MB13} is given as follows: 
\begin{align}
 &\theta_1 = \theta_3 = \pi,\ \theta_2 = 2\pi,
 \theta_4 = \theta_6 + 2\pi = 2\pi + \theta/2 -k,\ 
 \theta_5 = 2\pi -2k,\nonumber\\
 &\phi_1 = \phi_3 = \phi + \arccos[-\theta/(4\pi)],\ 
 \phi_2 = 3\phi_1 - 2\phi,
 \phi_4 = \phi_5 - \pi = \phi_6 = \phi,\ k = \arcsin[\sin(\theta/2)/2].
\end{align}

Table~\ref{table:pulse} shows parameters of $\pi/2$- 
and $\pi$-pulses of the above three composite pulses.

\subsection*{300~mM $^{13}$C-labelled chloroform in acetone-$d_6$}
$^{13}$C-labelled chloroform was purchased from Cambridge Isotopes.  
To the 300~mM $^{13}$C-labelled chloroform acetone-$d_6$ solution, $4$~mM of iron(III) 
acetylacetonate was added. Resulting $T_1(\rm^{13}C)$ and $T_2(\rm^{13}C)$ were $\sim 6$~s 
and $200$~ms, respectively, while $T_1(\rm^1H)$ and $T_2(\rm^1H)$ were both $\sim$ $200$~ms.

\subsection*{2\% HDO in D$_2$O}
To the solvent-mixture composed of 594~$\mu$L of D$_2$O and 6~$\mu$L of H$_2$O, 
2~mg of CuCl$_2$ was added, resulting in $T_1$($\rm^1H$) and $T_2$($\rm^1H$) 
of $\sim$50~ms at 9.7~T. 
Note that the solvent mixing causes 2~\% HDO solution, due to the H-D chemical exchange.

\subsection*{300~mM 3-chloro-2,4,5,6-tetrafluoro-benzotrifluoride in benzene-$d_6$}
3-chloro-2,4,5,6-tetrafluoro-benzotrifluoride was diluted 
with benzene-$d_6$ to 300~mM solution. 

\subsection*{NMR Measurements}
All the NMR experiments described in this article were measured on a JNM-ECA500 spectrometer 
(working at 11.7~T) 
or JNM-ECZ400S spectrometers (working at 9.4~T) (JEOL RESONANCE Inc.).  
The 2\% HDO sample was measured at 9.4~T (400 MHz for $\rm^1H$), 
and the other samples at 11.7~T (500 MHz for $\rm^1H$ and 488 MHz for $\rm^{19}F$). 
The measurements were carried out at 25{}$^\circ$C (9.4~T). 
A 5~mm (\{$\rm ^1H$, $\rm ^{19}F$\}-X) broadband (BB)  probe was used (11.7~T), 
and 5~mm ROYAL probes were used (9.4~T). 
We took $1/4$ of the square $2\pi$-pulse 
duration as the pulse duration of a square $\pi$/2-pulse (11.7~T). 
Instead, the nonlinear least square curve fitting method~\citep{kurimoto} was used (9.4~T).
During the $\rm^{13}C$ observing experiments, $\rm^1H$ are decoupled by WALTZ16 
decoupling trains. The acquired 2D time-domain data were processed as follows. 
For both $t_1$ and $t_2$ periods, the shifted sine-bell window function was 
multiplied. For $t_1$, zero-filling was done once. These data were then 
Fourier-transformed.

\begin{acknowledgements}
The authors acknowledge valuable discussions with Masahiro Kitagawa and Makoto Negoro. M. B. thanks the Institute for Molecular Science and the National Institutes of Natural Sciences for their hospitality. This work is partially supported by JSPS KAKENHI, Grant Numbers 24320008, 25400422, 25800181, 26400422, 16K05492, 17K05082, and 19K14636, the DAIKO Foundation, the Collaborative Research Project of the Laboratory for Materials and Structures, the Institute of Innovative Research, Tokyo Institute of Technology, the Joint Studies Program of the Institute for Molecular Science, and JST CREST, Grant Number JPMJCR1774.
\end{acknowledgements}


\begin{thebibliography}{99}
\bibitem{TDWC99} Claridge, T. D. W. {\it High-Resolution NMR Techniques in Organic Chemistry}, 3rd Edition (Elsevier Science, Amsterdam, 2016).

\bibitem{levitt} Levitt, M. H. {\it Spin Dynamics: Basics of Nuclear Magnetic Resonance}, 2nd Edition (John Wiley and Sons, New York, 2013).

\bibitem{JAJ11} Jones, J. A. Quantum computing with NMR. {\it Prog. Nucl. Magn. Reson. Spectrosc.} {\bf 59(2)}, 91--120 (2011).

\bibitem{CC85} Counsell, C., Levitt, M. H. \& Ernst, R. R. Analytical theory of composite pulses. {\it J. Magn. Reson.} {\bf 63(1)}, 133--141 (1985).
 
\bibitem{RT85} Tycko, R., Pines, R. A. \& Guckenheimer, J. Fixed point theory of iterative excitation schemes in NMR. {\it J. Chem. Phys.} {\bf 83}, 2775--2802 (1985).
 
\bibitem{MHL86} Levitt, M. H. Composite pulses. {\it Prog. Nucl. Magn. Reson. Spectrosc.} {\bf 18(2)}, 61--122 (1986).

\bibitem{MHL96} Levitt, M. H. in {\it Encyclopedia of nuclear magnetic resonance}, (eds. Grant D. M. \& Harris, R. K.)(Wiley, 1996).

\bibitem{ota} Ota, Y. \& Kondo, Y. Composite pulses in NMR as nonadiabatic geometric quantum gates. {\it Phys. Rev. A} {\bf 80}, 024302 (2009).

\bibitem{kindai} Ichikawa, T., Bando, M. Kondo, Y. \& Nakahara, M. Geometric aspects of composite pulses. {\it Phil. Trans. R. Soc. A} {\bf 370}, 4671 (2012). 

\bibitem{Elham} Lapasar, E. H., Maruyama, K.,  Burgarth, D., Takui, T., Kondo, Y. \& Nakahara, M. Estimation of coupling constants of a three-spin chain: a case study of Hamiltonian tomography with nuclear magnetic resonance. {\it New J. Phys.} {\bf 14}, 013043 (2012).
 
\bibitem{SW94} Wimperis, S. Broadband, narrowband, and passband composite pulses for use in advanced NMR experiments. {\it J. Magn. Reson. A} {\bf 109(2)}, 221--231 (1994).

\bibitem{HKC03} Cummins, H. K., Llewellyn, G. \& Jones, J. A. Tackling systematic errors in quantum logic gates with composite rotations. {\it Phys. Rev. A} {\bf 67}, 042308 (2003).
 
\bibitem{KRB04} Brown, K. R., Harrow, A. W., \& Chuang, I. L. Arbitrarily accurate composite pulse sequences. {\it Phys. Rev. A} {\bf 70}, 052318 (2004) (Errata {\bf 72}, 039905 (2005)).
 
\bibitem{WGA07} Alway W. G. \& Jones, J. A. Arbitrary precision composite pulses for NMR quantum computing. {\it J. Magn. Reson.} {\bf 189(1)}, 114--120 (2007).

\bibitem{MB13} Bando, M., Ichikawa, T., Kondo, Y. \& Nakahara, M. Concatenated composite pulses compensating simultaneous systematic errors. {\it J. Phys. Soc. Jpn.} {\bf 82}, 014004 (2013).

\bibitem{TI11} Ichikawa, T., Bando, M., Kondo, Y., \& Nakahara, M. Designing robust unitary gates: application to concatenated composite pulses. {\it Phys. Rev. A} {\bf 84}, 062311 (2011).

\bibitem{JAJ13} Jones, J. A. Designing short robust not gates for quantum computation. {\it Phys. Rev. A} {\bf 87}, 052317 (2013).
 
\bibitem{JTM12} Merrill, J. T. \& Brown, K. R. in {\it Quantum Information and Computation for Chemistry: Advances in Chemical Physics}, Volume 154, Ed. S. Kais, (John Wiley and Sons, Inc., Hoboken, New Jersey, 2014) p. 241.

\bibitem{Nielsen} Nielsen, M. A. \& Chuang, I. C. {\it Quantum Information and Quantum Computation}, (Cambridge University Press, Cambridge, 2000).

\bibitem{SZ} Suzuki, M. General theory of higher-order decomposition of exponential operators and symplectic integrators. {\it Phys. Lett. A} {\bf 165}, 387--395 (1992).

\bibitem{AJS83} Shaka, A. J., Keeler, J. \& Freeman, R. Evaluation of a new broadband decoupling sequence: WALTZ-16. {\it J. Magn. Reson.} {\bf 53(2)}, 313--340 (1983).

\bibitem{com} Siminovitch, D. J., Raleigh, D. P., Olejniczak, E. T. \& Griffin, R. G. Composite pulse excitation in quadrupole echo spectroscopy. {\it J.\ Chem.\ Phys.} {\bf 84}, 2556 (1986).

\bibitem{oc1} Palao, J. P. \& Kosloff, R. Optimal control theory for unitary transformations. {\it Phys. Rev. A} {\bf 68}, 062308 (2003) (Errata {\bf 69}, 059901 (2004)).

\bibitem{oc2} Khaneja, N., Reiss, T., Kehlet, C., Schulte-Herbr\"uggen, T. \& Glaser, S. J. Optimal control of coupled spin dynamics: design of NMR pulse sequences by gradient ascent algorithms. {\it J. Magn. Reson.} {\bf 172(2)}, 296--305 (2005).

\bibitem{oc3} Dong, D., \& Petersen, I. R. Quantum control theory and applications: a survey. {\it IET Control Theory and Applications} {\bf 4(12)}, 2651--2671 (2010).
  
\bibitem{Koike} Koike, T. \& Okudaira, Y. Time complexity and gate complexity. 
{\it Phys. Rev. A} {\bf 82}, 042305 (2010).

\bibitem{harvard} Novitski, E. M. {\it Apparatus and Methods for a New Measurement of the Electron and Positron Magnetic Moments}, Ph.D. Thesis, Harvard University (2017).

\bibitem{Mukamel1995} Mukamel, S. {\it Principles of Nonlinear Optical Spectroscopy} (Oxford University Press, New York, 1995).

\bibitem{kurimoto} Kurimoto, T., Asakura, K., Yamasaki, C. \& Nemoto, N. MUSASHI: NMR pulse width determination method by nonlinear least square curve fitting. {\it Chem. Lett.} {\bf 34(4)}, 540--541 (2005).
\end{thebibliography}
\end{document}